\documentclass[prl, twocolumn, showpacs, floatfix]{revtex4-1}
\usepackage{amssymb}
\usepackage{amsfonts}
\usepackage{amsmath}
\usepackage{graphicx}
\usepackage{enumerate}
\usepackage[usenames,dvipsnames]{color}
\usepackage{ulem}

\begin{document}

\title{A dynamical theory of homogeneous nucleation for colloids and
macromolecules}
\author{James F. Lutsko}
\affiliation{Center for Nonlinear Phenomena and Complex Systems, Code Postal 231,
Universit\'{e} Libre de Bruxelles, Blvd. du Triomphe, 1050 Brussels, Belgium}
\email{jlutsko@ulb.ac.be}
\homepage{http://www.lutsko.com}

\begin{abstract}
Homogeneous nucleation is formulated within the context of fluctuating
hydrodynamics. It is shown that for a colloidal system in
the strong damping limit the most likely path for nucleation can be
determined by gradient descent in density space governed by a nontrivial
metric. This is illustrated by application to low-density/high-density liquid transition of globular proteins in solution where it is shown that nucleation process involves two stages: the formation of an extended region with enhanced density followed by the formation of a cluster within this region.. 
\end{abstract}

\date{\today }
\maketitle

\paragraph*{Introduction}

Homogeneous nucleation is a fundamental physical process of importance in
fields as diverse as chemistry, materials science, biology and cosmology.
Our basic understanding of it goes back to Gibbs\cite{Kashchiev}.
The physics is governed by the fact that the excess free energy of a liquid
cluster relative to the vapor has a negative contribution that scales as the
volume and a positive contribution due to surface tension that scales as the surface area.
In Classical Nucleation Theory (CNT) it is assumed that the cluster
is spherical, that its interior is in the bulk liquid state and that the
surface tension is the same as for the coexisting liquid and vapor so that
the free energy of the cluster as a function of its radius can be calculated
giving a quantitative picture of homogeneous nucleation\cite{Kashchiev}.

This description has several shortcomings. The surface tension is generally
not constant
and, especially for small clusters, can
depend strongly on the size of the cluster. There is also no reason that the
density within the cluster should be constant or, even if it were, that it
should be equal to the bulk density for all cluster sizes. 
A more detailed picture can be devloped using  Density\ Functional Theory (DFT)\ from which provides models of the free
energy as a functional of the density profile(see e.g. Ref. \cite%
{Evans79,OxtobyEvans}). One can describe the density according to some
parametrization (such as a hyperbolic tangent) which will involve at least
three parameters:\ the central density, the radius and the interfacial
width, and proceed by choosing a reaction coordinate - such as the radius of
the cluster - and minimizing the free energy while holding the reaction
coordinate constant. This does indeed lead to finite interfacial widths and
size-dependent central densities, as expected(for recent examples, see Refs.%
\cite{Ghosh,Lutsko2011}).

Despite being physically reasonable, there are  significant
conceptual problems with this approach such as the arbitrarity of the reaction coordinate.
As described, the nucleation pathway will
consist of a monotonically increasing radius with the other parameters
determined by the minimization. However, one could just as well choose the
number of molecules in the cluster as the reaction coordinate in which case
- in principle - the radius need not increase monotonically along the
pathway since the mass of the cluster can increase by increasing the width
while, at the same time, decreasing the radius\cite{Lutsko2011}. Even the
excess mass is not a good reaction coordinate in general  and imposing more complex
constraints can, at least for some models, lead to spurious divergences\cite%
{LutskoBubble1, *LutskoBubble2}. 

The fundamental difficulty underlying 
these and other equilibrium, free-energy
based approaches is that the physical description is incomplete since
homogeneous nucleation is a fundamentally nonequilibrium, fluctuation-driven
process. This raises several questions about the classical description such
as whether the free energy plays such a central role and whether it is
necessary that the growing cluster actually pass through the critical
cluster. The solution is to develop a nonequilibrium, dynamical description
of homogeneous nucleation and this is the goal here. The following
development is based on Brownian Dynamics wherein molecules move
according to Newton's laws while being subject to a frictional force as well
as fluctuating forces. This is a simple model for colloids and the important
case of macromolecules in solution in which case the friction and the
fluctuations come from the bath/solvent.

\paragraph*{Theory}

The system consists of a collection of molecules of unit mass with positions
and momenta $\mathbf{q}_{i}$,$\mathbf{p}_{i}$ interacting via a potential $V$%
. Additionally, the particles interact with a bath/solvent of light
particles and this is described via a frictional drag and a fluctuating force%
\begin{equation}
\overset{\cdot }{\mathbf{q}}_{i}=\mathbf{p}_{i},\;\;\overset{\cdot }{\mathbf{%
p}}_{i}=-\frac{\partial V}{\partial \mathbf{q}_{i}}-\gamma \mathbf{p}_{i}+%
\mathbf{f}_{i}\left( t\right)
\end{equation}%
where all components of the fluctuating force are Gaussian and independent,%
\begin{equation}
\left\langle \mathbf{f}_{i}\left( t\right) \mathbf{f}_{j}\left( t^{\prime
}\right) \right\rangle =2\gamma k_{B}T\mathbf{1}\delta _{ij}\delta \left(
t-t^{\prime }\right)
\end{equation}
Defining the local density and momentum density respectively as%
\begin{equation}
\left\{ 
\begin{array}{c}
\widehat{\rho }\left( \mathbf{r};t\right) \\ 
\widehat{\mathbf{j}}\left( \mathbf{r};t\right)%
\end{array}%
\right\} =\sum_{i}\left\{ 
\begin{array}{c}
1 \\ 
\mathbf{p}_{i}%
\end{array}%
\right\} \delta \left( \mathbf{r}-\mathbf{q}_{i}\right)
\end{equation}%
one sees that these satisfy the exact equations%
\begin{align}
\frac{\partial \widehat{\rho }\left( \mathbf{r};t\right) }{\partial t}& =-%
\mathbf{\nabla \cdot }\widehat{\mathbf{j}}\left( \mathbf{r};t\right) \\
\frac{\partial \widehat{\mathbf{j}}\left( \mathbf{r};t\right) }{\partial t}&
=-\mathbf{\nabla \cdot }\sum_{i}\mathbf{p}_{i}\mathbf{p}_{i}\delta \left( 
\mathbf{r}-\mathbf{q}_{i}\right) -\sum_{i}\frac{\partial V}{\partial \mathbf{%
q}_{i}}\delta \left( \mathbf{r}-\mathbf{q}_{i}\right)  \notag \\
& -\gamma \widehat{\mathbf{j}}\left( \mathbf{r};t\right) +\sqrt{2\gamma
k_{B}T\widehat{\rho }\left( \mathbf{r};t\right) }\mathbf{F}\left( \mathbf{r;}%
t\right)  \notag
\end{align}%
with%
\begin{equation}
\left\langle \mathbf{F}\left( \mathbf{r;}t\right) \mathbf{F}\left( \mathbf{r}%
^{\prime }\mathbf{;}t^{\prime }\right) \right\rangle =\delta \left(
t-t^{\prime }\right) \delta \left( \mathbf{r}-\mathbf{r}^{\prime }\right) 
\mathbf{1}
\end{equation}%
Coarse graining in space and assuming local equilibrium leads to a
mesoscopic description of fluctuations in terms of fluctuating
hydrodynamics. Neglecting temperature fluctuations gives 
\begin{align} \label{hydro}
\frac{\partial \rho \left( \mathbf{r}\right) }{\partial t}+\mathbf{\nabla
\cdot j}\left( \mathbf{r}\right) =0 & \\
\frac{\partial \mathbf{j}\left( \mathbf{r}\right) }{\partial t}+\mathbf{%
\nabla \cdot j}\left( \mathbf{r}\right) \mathbf{j}\left( \mathbf{r}\right)
/\rho \left( \mathbf{r}\right) +\rho \left( \mathbf{r}\right) \mathbf{\nabla 
}\frac{\delta F\left[ \rho \right] }{\delta \rho \left( \mathbf{r}\right) } &
\notag \\
+\mathbf{\nabla \cdot \Pi }\left( \mathbf{r}\right) =-\gamma \mathbf{j}%
\left( \mathbf{r}\right) +\sqrt{2\gamma k_{B}T\rho \left( \mathbf{r}\right) }%
& \mathbf{\xi }\left( \mathbf{r};t\right)  \notag
\end{align}%
where $\rho \left( \mathbf{r}\right) $ and $\mathbf{j}\left( \mathbf{r}%
\right) $ are the coarse-grained local density and momentum density, $F\left[
\rho \right] $ is the coarse-grained free energy and $\mathbf{\Pi }$ is the
dissipative part of the stress tensor which has both a deterministic and a
fluctuating contribution\cite{Chavanis1, *Chavanis}. The free energy
term is a representation of the local pressure and has been discussed
extensively in the DFT literature
: its use here can be viewed as a local equilibrium approximation\cite{Dean, Evans79, EvansArcher, Archer}. The quantity $\mathbf{\xi}\left( \mathbf{r};t\right)$ is the noise due to
the Brownian dynamics and is white and delta-correlated in space and time.
Note that this is just the natural generalization of Landau and Lifshitz's
fluctuating hydrodynamics taking account of the Brownian forces. Assuming
that the velocity will always be small due to the damping, the convective
term can be neglected so that the second equation becomes linear in the
momentum density. Eliminating the momentum current then gives 
\begin{align}
\frac{\partial ^{2}\rho \left( \mathbf{r}\right) }{\partial t^{2}}+\gamma 
\frac{\partial \rho \left( \mathbf{r}\right) }{\partial t}-&\mathbf{\nabla
\cdot }\left( \rho \left( \mathbf{r}\right) \mathbf{\nabla }\frac{\delta F%
\left[ \rho \right] }{\delta \rho \left( \mathbf{r}\right) }\right)
\label{OD} \\
+\mathbf{\nabla }\cdot \sqrt{2\gamma k_{B}T\rho (\mathbf{r})}\mathbf{\xi }%
\left( \mathbf{r};t\right) & =0  \notag
\end{align}%
In the following, the second-time derivative, the so-called inertial term,
will be neglected, as is usual in the strong-damping approximation. Then,
when the density is low, in the ideal gas limit, the first term on the right
becomes $\gamma ^{-1}k_{B}T\nabla ^{2}\rho \left( \mathbf{r}\right) $ so
that $D\equiv \gamma ^{-1}k_{B}T$ can be identified as the diffusion
constant.

The use of fluctuating hydrodynamics as basis for studying nucleation is
similar to the approach developed by Langer\cite{Langer1,*Langer2}. The
primary difference here is that the emphasis is on understanding the
time-evolution of the formation of the critical cluster whereas previous
work focused on the nucleation rate. This development differs from more
phenomenological approaches which are couched entirely in terms of order
parameters, such based on nonequilibrium thermodynamics\cite{Nicolis} or
phase field theory\cite{Gunton}, in that nonlinearities of the transport
coefficients and colored noise occur naturally and play an important role.
One of the goals below is to relate the hydrodynamic description to one
involving order parameters.

In order to characterize the generic properties of the process of
nucleation, we focus here on the \emph{most likely path} (MLP) where a
\textquotedblleft path\textquotedblright\ is understood as a function $\rho
(\mathbf{r};t)$ connecting the initial state of pure metastable phase and the final
state of pure stable phase. When the noise amplitude is small (as in the
strong damping limit), most systems should go through a nucleation pathway
close to this generic result. 
In general, determining the MLP is complex. However, without the inertial
term, Eq.(\ref{OD}) is a gradient-driven, diffusive dynamics which obeys a
fluctuation-dissipation relation. By a straightforward generalization of 
\cite{Heymann}, it can be shown that \emph{for this type of dynamics the MLP
connecting metastable states does indeed pass through the critical point} and
that it coincides with either the forward-time or backward-time
deterministic trajectory in density space\cite{supp}. The MLP can therefore
be determined by starting at a local minimum and moving along the
deterministic path 
\begin{equation}
\frac{\partial \rho \left( \mathbf{r}\right) }{\partial t}=\pm D\mathbf{%
\nabla \cdot }\left( \rho \left( \mathbf{r}\right) \mathbf{\nabla }\frac{%
\delta \beta F\left[ \rho \right] }{\delta \rho \left( \mathbf{r}\right) }%
\right)   \label{rad}
\end{equation}%
where the sign is chosen according to the direction one wishes to move\cite{Heymann,supp}. 

Equation(\ref{rad}) is
the primary theoretical result of this paper. It superficially resembles the
usual Dynamic Density Functional Theory (DDFT) equation\cite{Evans79, EvansArcher, Archer, Chavanis} but
is in fact considerably more general. It says that the most likely path can
be determined by following the DDFT dynamics \emph{when that dynamics does
indeed connect the desired states} such as in passing from a high-energy to
a low-energy state with no barrier separating them. (An example of this
would be spinodal decomposition\cite{EvansArcher}.) However, DDFT
cannot describe the crossing of a free energy barrier as it specifically
pertains to the ensemble-averaged density. In essence, it is the result of
averaging Eq.(\ref{OD}) (without the inertial term) over the noise. In
contrast, Eq.\ (\ref{rad})  also describes the MLP when this means going
uphill against the free energy gradient. In that case, it says that the MLP
can be obtained by reversing the sign of the gradient or, equivalently, by
following the time-reversed dynamics\cite{Heymann,supp}. It can
therefore be viewed as an extension of  DDFT to barrier-crossing problems,
given the various assumptions set out above. This simple result is strongly
dependent on the existence of the fluctuation-dissipation relation in Eq.(%
\ref{OD}) and on the assumption of weak noise (compared to the thermodynamic
driving force). It will not be exact if either of these conditions are violated
and, in particular, the much more complicated strong-noise result will be
discussed in at a later time. Finally, it is important to note that Eq.(\ref%
{rad})\ is simply a mathematical means of identifying the MLP and that it
does not imply in any sense that the actual (strongly-dissipitive) dynamics
is time-reversal invariant.  

In principle, Eq.(\ref{rad}) could be integrated directly to determine the
MLP or some other technique, such as the string method, used to determine
the path. However, the goal here is to generalize previous descriptions of
nucleation which are based on a set of order parameters characterizing the
system. In CNT, the cluster is assumed to be spherical and only parameter is
the size of the cluster: more generally, a minimal set would include some
measure of the density inside the cluster and the width of the interface as
well. In the present formalism, the order parameters must somehow be related
to the spatial density since it is the fundamental quantity describing the
evolution. We therefore imagine that the density profile is approximated by
some test function of the form $\rho \left( r,t\right) =f\left( r;\mathbf{x}%
\left( t\right) \right) $ where $\mathbf{x}\left( t\right) $ stands for the
set of order parameters. It is possible to give an exact equation for the
evolution of the parameters based on an analysis of the MLP but here a more
heuristic method is used. First, Eq.(\ref{rad}) is integrated over a
spherical volume of radius $r$ giving%
\begin{equation}
\frac{\partial m\left( r;\mathbf{x}(t)\right) }{\partial t}=\pm D\int_{S\left( r\right)
}\rho \left( \mathbf{r^{\prime }}\right) \left( \frac{\partial }{\partial
r^{\prime }}\frac{\delta \beta F\left[ \rho \right] }{\delta \rho \left( 
\mathbf{r^{\prime }}\right) }\right) dS^{\prime }  \label{SS}
\end{equation}%
where $m(r;\mathbf{x}(t))$ is the mass in side the spherical shell of radius $r$ and the
notation indicates a surface integral over that shell. Then, spherical
symmetry is assumed and Eq.(\ref{SS}) is multiplied by $r^{-2}\rho
^{-1}\left( r\right) \frac{\partial m(r)}{\partial x_{a}}$ and integrated
over $r$ to get%
\begin{equation}
g_{ab}\frac{dx_{b}}{dt}=\pm D\frac{1}{4\pi }\frac{\partial \beta \Omega }{%
\partial x_{a}}  \label{grad}
\end{equation}%
where the metric is 
\begin{equation}
g_{ab}=4\pi \int_{0}^{\infty }\frac{1}{r^{2}\rho \left( r\right) }\frac{%
\partial m\left( r\right) }{\partial x_{a}}\frac{\partial m\left( r\right) }{%
\partial x_{b}}dr  \label{SD2}
\end{equation}%
and where $\Omega =F-\mu N$ is the grand potential which arises due to an
integration by parts. This becomes exact if the parametrization is complete
in the sense that $f\left( r;\mathbf{x}\left( t\right) \right) $ is able to
represent any well-behaved function arbitrarily closely(e.g. an expansion in
a complete set of basis functions). The exact minimization of the action for
the case of a finite number of order parameters and its relation to this
approximation will be discussed elsewhere.

\paragraph*{Application to low density/high density liquid transition in globular proteins}

Many proteins in solution exhibit a phase transition between a low density gas-like phase and a high density liquid-like phase. This behavior can be modeled using an effective pair-potential in which case it becomes analogous to the vapor/liquid transition in simple fluids. Calculations were performed for the
ten Wolde-Frenkel\cite{tWF}  model potential for globular proteins having hard-core diameter $\sigma$ and energy scale $%
\epsilon$ using the squared-gradient free energy model, 
\begin{equation}
F[\rho] = \int \left(f(\rho(\mathbf{r}))+\frac{1}{2}K(\nabla \rho(\mathbf{r}%
))^2 \right)d\mathbf{r}
\end{equation}
where $f(\rho )$ is the bulk free energy per unit volume, calculated using thermodynamic perturbation theory, and the coefficient $K$ was calculated using a recently-derived
approximation\cite{Lutsko2011}. Equation (\ref{rad}) was integrated (assuming spherical symmetry) by discretizing the right hand side using Eq.(\ref{grad}) and the method of piece-wise linear profiles\cite{Lutsko2011} (equivalent to a variable grid method). At a temperature of  $k_{B}T=0.375\epsilon $ and with a pressure supersaturation of $1.159$,  the exact excess energy barrier was found to be $\Delta
\beta \Omega = 75.8$ with $1158$ molecules in the cluster while the discretization with 19 parameters gives a value of $77.1$ and $1175$ molecules. The MLP was then determined by starting near the at the critical cluster and perturbing in the direction of the negative eigenvalue\cite{Wales} and
integrating Eq.(\ref{grad}) numerically\cite{Intel}. In tracing the backwards part of the path, the calculations were terminated when $\beta \Delta \Omega = 1k_BT$ since the weak noise approximation is not applicable for lower energies\cite{Bier}. 

\begin{figure}[tbp]
\includegraphics[angle=-0,scale=0.3]{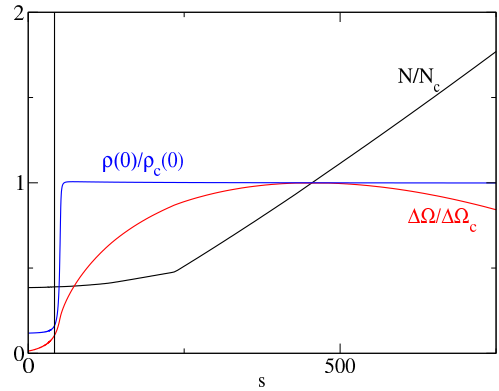} 
\caption{The excess particle number, excess free energy and central density
relative to their values in the critical cluster as functions of the natural reaction coordinate, $s$, where $s$=0 corresponds to the initial vapor phase and the final liquid phase occurs for $s=\infty$. The vertical line marks the transition between the two growth regimes (see text).}
\label{fig1}
\end{figure}

\begin{figure}[tbp]
\includegraphics[angle=-0,scale=0.3]{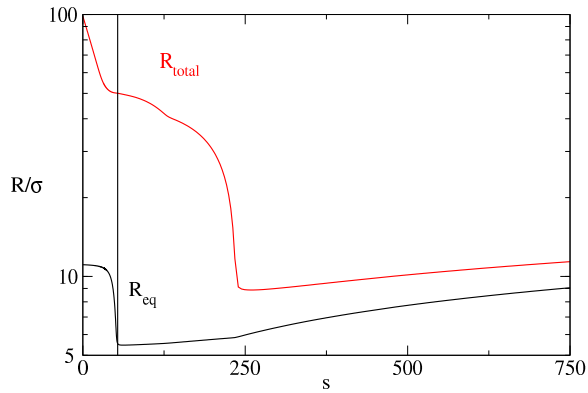} 
\caption{The same as Fig. \protect\ref{fig1}, but showing the equimolar
radius, $R_{eq}$, and the total spatial extent, $R_{total}$, along the path. The vertical line marks the same point as in Fig.\ref{fig1}.}
\label{fig2}
\end{figure}

Figure \ref{fig1} shows the evolution of the excess number of particles,
excess free energy and of the central density. The independent variable is
the natural reaction coordinate which is distance along the nucleation
pathway as calculated using the metric, Eq.(\ref{SD2}). When the cluster is
large, the path is similar to that which would be obtained using typical
heuristic methods. However, for smaller droplets, the results are quite
nonclassical. Figure \ref{fig2} gives the spatial size of the droplet
according to two different measures: the equimolar radius as calculated
based on the central density and the (model-dependent) total spatial extent
of the droplet (in this model, the droplet always has a well-defined finite
support). Combining the information in these two figures, it is seen that
the MLP begins with a spatially-extended disturbance having very low density
but a fixed excess number of molecules (in the present case, about 450).

The interpretation of these results is not as different from the usual picture of nucleation as they might at first appear. At short times, during which the equimolar radius is nearly constant,
a small increase in density forms over a spatially extended volume. From Fig. 1, it is apparant that, despite its spatial extent, the excess energy
of this density fluctuation is quite small so that its formation is not improbable. The second part of the process is the formation of a cluster
within this region of enhanced density. That nucleation would preferentially take place in a region of enhanced density, which therefore already contains 
some of the excess mass needed to form a cluster, is quite reasonable. Indeed, having the necessary excess mass present locally allows the cluster to form more rapidly than if matter had to diffuse in from the surrounding bulk. What does appear strange is the directed nature of process with the density fluctuation
 appearing to contract to form the nucleus. This is partly a result of insisting on spherical symmetry and partly due to a well-known property of the MLP wherein 
it typically involves a system crossing a barrier in the shortest number of steps possible with no back-tracking or variation\cite{Bier}. The MLP is of course an abstraction: 
an actual realization of the processes will involve fluctuations around it and will not appear so deterministic( see, e.g., Ref\cite{Bier}). Finally, because mass is conserved, \emph{any} dynamical process of cluster formation is going to give the appearance of drawing in mass from the boundaries of the system.

\paragraph{Conclusions}

A description of nucleation applicable to colloids and macromolecules in
solution based on fluctuating hydrodynamics has been developed. It was shown
that under assumptions of strong dissipation and weak noise the most likely path could be determined by gradient descent on the
free energy surface and that it necessarily passes through the critical point, thus providing justification for more heuristic methods
based solely on free energy considerations\cite{LutskoBubble1,*LutskoBubble2}. 
It is
also interesting to note that Eq.(\ref{grad}) can be seen to
justify more phenomenological treatments of nucleation in which a set of
order parameters is assumed to evolve stochastically as $dx/dt = L\frac{%
\delta F}{\delta x} + \xi$ with a fluctuation-dissipation relation
determining the amplitude of the noise. The same approach can be applied to other nucleation
phenomena such as heterogeneous nucleation, nucleation in confined systems and even, conceivably, to transitions in
granular fluids. 

\begin{acknowledgments}
It is a pleasure to thank Gr\'{e}goire Nicolis for several insightful
comments. This work was supported in part by the European Space Agency under
contract number ESA AO-2004-070.
\end{acknowledgments}

%

\end{document}